\documentclass[11pt,a4paper]{article}
\usepackage{jheppub}

\usepackage{amsmath,graphicx,amsfonts,mathrsfs,amssymb}
\usepackage{amsmath,epsfig}
\usepackage{amssymb,amsfonts}
\usepackage{latexsym}
\usepackage{epsfig}
\newbox\pippobox
\def\be{\begin{equation}}
\def\ee{\end{equation}}
\def\bea{\begin{eqnarray}}
\def\eea{\end{eqnarray}}

\def\ee           {{\rm e}}

\newcommand{\beq}{\begin{equation}}
\newcommand{\eeq}{\end{equation}}
\newcommand{\beqa}{\begin{eqnarray}}
\newcommand{\eeqa}{\end{eqnarray}}
\newcommand{\beqar}{\begin{eqnarray*}}
\newcommand{\eeqar}{\end{eqnarray*}}

\renewcommand{\eqref}[1]{(\ref{#1})}

\catcode`\@=12

\title{A hQCD model and its phase diagram in Einstein-Maxwell-Dilaton
system}
\author[a,b]{Rong-Gen Cai}
\author[a,b]{Song He,}
\author[c]{Danning Li,}

\affiliation[a]{State Key Laboratory of Theoretical Physics,
Institute of Theoretical Physics, Chinese Academy of Science,
Beijing 100190, People's Republic of China } \affiliation[b]{ Kavli
Institute for Theoretical Physics China, CAS, Beijing 100190, China}
\affiliation[c]{Institute of High Energy Physics, Chinese Academy of
Sciences, Beijing, China}

\emailAdd{cairg@itp.ac.cn}\emailAdd{hesong@itp.ac.cn}\emailAdd{lidn@ihep.ac.cn}

\date{\today}

\abstract{By use of  the potential reconstruction approach we obtain
 a series of asymptotically AdS (aAdS) black hole
solutions in an Einstein-Maxwell-Dilaton (EMD) system. Basing on the
solutions of the system, we reconstruct a semi-analytical
holographic QCD (hQCD) model with a quadratic term  in warped
factor. We discuss some aspects of the hQCD model, in particular we
calculate the free energy of two static color sources (a heavy
quark-antiquark pair) which is an important order parameter to
describe confinement/deconfinement phase transition. The behavior of
the free energy with respect to temperature and chemical potential
is studied. We find that in the hQCD model the deconfinement phase
transition can be realized and a critical point occurs.  The
resulting phase diagram in the temperature-chemical potential
$T-\mu$ plane is in quite good agreement with the one from recent
lattice results and effective models of QCD.}


\begin{document}

\maketitle

\section{Introduction}
Understanding strongly coupling QCD related to the RHIC and LHC
experiments is attracting much attention. Although a powerful method
for this subject, the lattice QCD, is being developed, when it comes
to the famous sign problem, lattice techniques are not well adapted
to the case with finite real chemical potential $\mu$ and no many
results have been produced so far. The other effective approaches,
models and effective field theories also suffer from some problems
in the dense matter case. On the other hand, the gauge/gravity
duality
\cite{Maldacena:1997re,Gubser:1998bc,Witten:1998qj,Aharony:1999ti}
developed in string theory, can offer new insights to hadron theory
\cite{Erlich:2005qh,TB:05,DaRold2005,D3-D7,D4-D6,SS,Dp-Dq,Csaki:2006ji}
and strongly interacting quark gluon plasma (QGP)
\cite{Shuryak:2004cy,Policastro:2001yc,Cai:2009zv,Cai:2008ph,Sin:2004yx,Nastase:2005rp,
Janik:2005zt,Herzog:2006gh} in top-down and bottom-up setups. The
duality approach can easily deal with the dense matter problem at
least for the case in the deconfinement phase. However, for the
hadron phase, the current status is still not very satisfied because
the builded phase diagram so
far\cite{Rozali:2007rx,DeWolfe:2010he,Horigome:2006xu,Bergman:2007wp}
 is different from that of the real QCD
~\cite{Kogut:2004su,Stephanov:2007fk,Alford:2007xm}. One is
expecting to build up a holographic model which can completely
recover the phase diagram of the real QCD.

 If one considers  a system with
graviton and
dilaton~\cite{Gursoy:2007cb,Gursoy:2007er,Gubser-T,Gursoy-T,Megias:2010ku},
one can obtain solutions of  Einstein equations by using the
potential reconstruction method proposed in
\cite{Li:2011hp,He:2011hw}. However, in previous works, the back
reacted geometry of the hadronic phase is not fully identified: if
one wants to consider the degrees of freedom of quarks, one should
take $U(1)$ gauge field into consideration in the
bulk~\cite{Rozali:2007rx,Charmousis:2010zz,Gouteraux:2011ce,DeWolfe:2010he,DeWolfe:2011ts,Evans:2011mu,Evans:2011tk,Evans:2011eu,Andreev:2010bv}.
Adding the additional field to the graviton-dilaton system, the
$U(1)$ gauge field is dual to the baryon number current
$J_D=\,\psi^\dagger(x) \psi(x)$, one may generate a chemical
potential by turning on an appropriate electric field in the black
hole geometry. On the other hand, in QCD the effect of finite quark
density is introduced by adding the term $J_D=\mu\,\psi^\dagger(x)
\psi(x)$ to the Lagrangian in the generating functional, so that the
chemical potential $\mu$ appears as the source of the quark density
operator. According to $AdS$/CFT correspondence, the source of a QCD
operator in the generating functional is the boundary value of a
dual field in the bulk; therefore, the chemical potential can be
considered as the boundary value of the time component of a $U(1)$
gauge field $A_M$ dual to the vector quark current. Note that the
$U(1)$ denotes the gauge symmetry in the bulk and global symmetry on
the boundary. The symmetry is held by physical requirement of
preservation of baryon number. Motivated from the holographic
description of unquenched QCD system, we will design a
graviton-dilaton-U(1) gauge field system to accommodate the degrees
of freedom in full QCD.

The confinement/deconfinement phase transition in QCD phase diagram
is a very interesting issue to play with holography. In holographic
QCD, it has been widely believed that the confinement phase in the
pure Yang-Mills theory corresponds to the AdS $D4$ soliton in
gravity and the deconfinement phase corresponds to the black D4
brane pointed out by \cite{Witten:1998zw}. \cite{hep-th/0608151}
argued that deconfinement in hard wall and soft wall models occurs
via a first order Hawking-Page type phase transition between a low
temperature thermal AdS space and a high temperature black hole.
\cite{Cai:2007zw} extended this discussion by studying (charged) AdS
black holes with spherical or negative constant curvature horizon.
The authors of \cite{Witten:1998zw} and \cite{Lee:2009bya}
investigated the deconfinement phase transition by introducing hard
wall in the AdS/Reissner-Nordstr\"om black-hole with using
Hawking-Page phase transition. Recently, the authors of
\cite{arXiv:1107.4048} and \cite{arXiv:1111.5190} carefully
considered the correspondence between phases and gravity backgrounds
proposed by \cite{Witten:1998zw} and argued that the alternative
gravitational configuration named ¡°localized soliton¡± would be
properly related to the deconfinement phase. The deconfinement
transition can be realized as a Gregory-Laflamme type transition.
There are also various studies on Polyakov loop
\cite{Zeng:2008sx,Andreev-T1,Andreev-T2,Andreev-T3,Noronha-T} to
investigate  QCD phase structure within hQCD models. We will propose
a hQCD model and  study its phase structure by studying the behavior
of Polyakov loop operator in this model.

In this paper, we will extend  the potential reconstruction approach
\cite{Li:2011hp,He:2011hw} to an Einstein-Maxwell-Dilaton (EMD)
system and build up a hQCD model with a positive quadratic term in
warped factor of the bulk metric. Furthermore, we will calculate the
free energy of two connecting Polyakov loop operators from
holographic point of view to check whether the holographic model can
realize the deconfinement phase transition. We introduce the
quadratic correction term in warped factor to construct the gravity
configuration with asymptotic AdS UV behavior. The motivation of
choosing the quadratic correction is that the quadratic correction
plays important roles to realize various facts in low energy QCD.
The work \cite{Karch:2006pv} by introducing the dilaton with form of
$e^{-c^2 z^2}$ can realize the Regge behavior of hadron spectrum
which can not be achieved in hard wall hQCD
model\cite{Erlich:2005qh}\cite{Polchinski:2002jw}. Andreev and
Zakharov introduced a positive quadratic correction, $e^{cz^2}$ with
$z$ the holographic coordinate and $c>0$, to the warp factor of
${\rm AdS}_5$ geometry. It turns out it is helpful to realize the
linear behavior in heavy quark potential \cite{Andreev:2006ct}. The
linear heavy quark potential can also be obtained by introducing
other deformed warp factors, e.g. the deformed warp factor which
mimics the QCD running coupling \cite{Pirner:2009gr}, and the
logarithmic correction with an explicit IR cutoff $\log
\frac{z_{IR}-z}{z_{IR}}$ \cite{He:2010ye}. To produce the linear
Regge behavior of the hadron excitations, Karch, Katz, Son and
Stephanov \cite{Karch:2006pv} proposed the soft-wall ${\rm AdS}_5$
model or KKSS model by introducing a quadratic correction to dilaton
background in the 5D meson action, whose effect in some sense looks
like introducing a negative quadratic correction, $e^{-cz^2}$ in the
warp factor of the ${\rm AdS}_5$ geometry. However, it is worth
mentioning that the model with a quadratic correction in the warp
factor of the metric is not equivalent to the model with a quadratic
correction in the dilaton background, although both models give the
same effective potential of hadron spectrum in the IR region~
\cite{Gursoy:2007cb,Gursoy:2007er}. A positive quadratic correction
$e^{cz^2}$ in the dilaton background of the 5D hadron action has
also been used to investigate hadron spectra
\cite{Zuo:2009dz,deTeramond:2009xk}, however, higher spin
excitations in this background will lead to imaginary mass
\cite{Dp-Dq,KKSS-2} and they also are inconsistent in the gluon
sector. There are the flavor models
\cite{Casero:2007ae,Iatrakis:2010zf,Iatrakis:2010jb} taking into
account the chiral condensation. In \cite{Sui:2009xe,Sui:2010ay},
the modified five-dimensional metric at the infrared region is
constructed to obtain a nontrivial dilaton solution, which
incorporates the chiral symmetry breaking and linear confinement.
The modified factor is leading terms $1+k^2 z^2$ in $e^{k^2 z^2}$.
They do predict the mass spectra of resonance states in the
pseudoscalar, scalar, vector and axial-vector mesons, which agree
with experiment data. \cite{Zhang:2010tk} also confirmed that the
quadratic term in dilaton plays an important role in linear
confinement in the meson sector. Recently, basd on the AdS
Reissner-Nordstr\"om black hole metric  the work
\cite{Colangelo:2010pe} introduced the $e^{k^2 z^2}$ term in the
dilaton, it turns out to be helpful to realize the heavy quark
potential. All these studies mentioned above do not consider effects
of back reaction of dilation and/or a modified warped factor . The
work \cite{He:2010ye} studied the topic in a non-critical string
framework with back reaction effects and found that the quadratic
term $e^{k^2 z^2}$ in warped factor is helpful to achieve the
Cornell potential. The work \cite{Li:2011hp} further confirmed the
conclusion from thermal hQCD perspective. In some sense, a positive
quadratic term captures some important QCD features, although we
still do not well understand. Therefore we choose the warped factor
with a positive quadratic correction to build up an effective hQCD
model.

The organization of the paper is as follows. In Section 2 we briefly
introduce the potential reconstruction approach  and apply it to an
Einstein-Maxwell-Dilaton system (EMD). In Section 3, we impose
asymptotical AdS boundary conditions and regularity requirements on
generic black hole solutions and figure out general formulas of
thermodynamic quantities of the black holes. In Sections 4 , we make
use of the reconstruction approach to propose our hQCD model with a
positive quadratic $k^2 z^2$ term in warped factor and calculate
relevant thermodynamic quantities such as temperature, entropy, etc,
of the background solution. We also simply check the stability of
dilaton potential from AdS/CFT perspective. In Section 5, we
investigate free energy of a heavy quark pair and vacuum expetation
value (VEV) of Polyakov loop which is order parameter to describe
the deconfinement phase transition. In this hQCD model, the
deconfinement phase transition can be realized and a critical point
appears. Section 6 is devoted to conclusions and discussions.

\section{Einstein-Maxwell-Dilaton  system}
\label{gravitysetup}

Let us begin with the following 5D Einstein-Maxwell-Dilaton (EMD)
action in string frame
\begin{equation} \label{minimal-String-action}
S_{5D}=\frac{1}{16 \pi G_5}\int d^5 x \sqrt{-g^S} e^{-2 \phi}
 \left(R^S + 4\partial_\mu \phi
\partial^\mu \phi-
V_S(\phi)-\frac{1}{4g_{g}^2}e^{\frac{4\phi}{3}}F_{\mu\nu}F^{\mu\nu}\right),
\end{equation}
where $G_5$ and $g_g$ are 5D Newtonian constant and effective gauge
coupling constant, $g^S$ and $V_S(\phi)$ are the 5D metric and
dilaton potential in the string frame, respectively.

To solve this system, it is convenient to transform this action to
Einstein frame. If the metric in  Einstein frame $g^E_{\mu\nu}$ and
its corresponding one in  string frame $g^S_{\mu\nu}$ are connected
by the scaling transformation
\begin{equation}
g^S_{\mu\nu} = e^{4 \phi \over 3 }g^E_{\mu\nu},
\end{equation}
one can derive the exact relation between two actions in string
frame and Einstein frame
\begin{eqnarray}\label{string-enssteinframe}
& & \int \sqrt{-g^S} e^{-2 \phi} \left(R^S + 4\partial_\mu \phi
\partial^\mu \phi- V_S(\phi)-\frac{1}{4g_{g}^2}e^{\frac{4\phi}{3}}F_{\mu\nu}F^{\mu\nu}\right) \nonumber \\
&=&\int \sqrt{-g^E} \left[ R^E -\frac{4}{3}
\partial_\mu \phi \partial^\mu \phi - V_E(\phi)-\frac{1}{4g_{g}^2}F_{\mu\nu}F^{\mu\nu}\right]
\end{eqnarray}
up to a total derivative term,  where \begin{equation} V_S=V_E
e^{\frac{-4\phi}{3}}.
\end{equation}
Thus we have the action in Einstein frame
\begin{equation} \label{minimal-Einstein-action}
S_{5D}=\frac{1}{16 \pi G_5} \int d^5 x
\sqrt{-g^E}\left(R-\frac{4}{3}\partial_{\mu}\phi\partial^{\mu}\phi-V_E(\phi)-\frac{1}{4g_{g}^2}F_{\mu\nu}F^{\mu\nu}
\right),
\end{equation}
where $F_{\mu\nu}=\partial_\mu A_\nu-\partial_\nu A_\mu$ is Maxwell
field.  Note that here we do not consider the coupling between gauge
field and dilaton field in Einstein frame.

For our aim, we are looking for black hole solutions of the system.
 In string frame we suppose the metric is of the form
\begin{equation} \label{metric-stringframe}
ds_S^2=\frac{L^2
e^{2A_s}}{z^2}\left(-f(z)dt^2+\frac{dz^2}{f(z)}+dx^{i}dx^{i}\right),
\end{equation}
with $L$ the radius of ${\rm AdS}_5$.  We will set $G_5=1$ and $g_g
L=1$ in Section 3, 4 and 5. In Einstein frame, the metric becomes
\begin{eqnarray} \label{metric-Einsteinframe}
ds_E^2=\frac{L^2 e^{2A_s-\frac{4\phi}{3}}}{z^2}\left(-f(z)dt^2
+\frac{dz^2}{f(z)}+dx^{i}dx^{i}\right).
\end{eqnarray}
In Einstein frame, the gravitational field equations read
\begin{eqnarray} \label{EOM}
E_{\mu\nu}+\frac{1}{2}g^E_{\mu\nu}\left(\frac{4}{3}
\partial_{\mu}\phi\partial^{\mu}\phi+V_E(\phi)\right)
-\frac{4}{3}\partial_{\mu}\phi\partial_{\nu}\phi -\frac{1}{2
g_{g}^2}\left(F_{\mu \lambda}F_\nu^\lambda-\frac{1}{4}
g^E_{\mu\nu}F_{\rho\sigma}F^{\rho\sigma}\right)=0,\nonumber\\
\end{eqnarray}
where $E_{\mu\nu}$ is Einstein tensor. In the metric
(\ref{metric-Einsteinframe}), the $(t,t), (z,z)$ and $(x_1, x_1)$
components of the gravitational field equations are respectively
 \bea &{}&b''(z)+\frac{b'(z) f'(z)}{2 f(z)}+\frac{2}{9}
b(z) \phi '(z)^2+\frac{b(z)^3 V_E(\phi (z))}{6
f(z)}+\frac{A_{t}'(z)^2}{24 g_g^2 b(z)
f(z)}=0\label{equation-graviton1},\\&{}& \phi '(z)^2-\frac{9
b'(z)^2}{b(z)^2}-\frac{9 b'(z) f'(z)}{4 b(z) f(z)}-\frac{3 b(z)^2
V_E(\phi (z))}{4 f(z)}-\frac{3 A_{t}'(z)^2}{16 g_g^2 b(z)^2
f(z)}=0\label{equation-graviton2},\\&{}&f''(z)+\frac{6 b'(z)
f'(z)}{b(z)}+\frac{6 f(z) b''(z)}{b(z)}+\frac{4}{3} f(z) \phi
'(z)^2+b(z)^2 V_E(\phi (z))-\frac{A_t'(z)^2}{4 g_g^2
b(z)^2}=0\nonumber,\\\label{equation-graviton3}\eea  where
$b(z)=\frac{L^2 e^{2A_E}}{z^2} $, $A_E(z)=A_s(z)-\frac{2}{3}\phi(z)$
and $A_t(z)$ is electrical potential of Maxwell field. As a
consistent check, turning off $A_t(z)$ in
Eqs.~(\ref{equation-graviton1}) -(\ref{equation-graviton3}), one can
easily reproduce equations of motion about the graviton-dilaton
system discussed in \cite{Li:2011hp}.

Note that the above three equations are not independent.  In
(\ref{equation-graviton1})-(\ref{equation-graviton3}), there are
only two independent functions, therefore one of the three equations
can be used to check the consistence of solutions. From those three
equations one can obtain following two equations which do not
concern the dilaton potential $V_E(\phi)$,
\begin{eqnarray}\label{AF}
 &{}&A_s''(z)+A_s'(z) \left(\frac{4 \phi '(z)}{3}+\frac{2}{z}\right)-A_s'(z)^2-\frac{2 \phi ''(z)}{3}-\frac{4 \phi '(z)}{3 z}=0\\\label{ff}
 &{}&f''(z)+ f'(z)\left(3 A_s'(z)-2 \phi '(z)-\frac{3 }{z}\right)-\frac{z^2 e^{\frac{4 \phi (z)}{3}-2 A_s(z)} A_t'(z){}^2}{ g_{g}^2 L^2}=0.
\end{eqnarray}
 Eq.(\ref{AF}) is our starting point
to find exact solutions of the system.  Note that Eq.(\ref{AF}) in
the EMD system is the same as the one in the Einstein-dilaton system
considered in \cite{Li:2011hp,He:2011hw} and the last term in
Eq.(\ref{ff}) is an additional contribution from electrical field.

The equation of motion of the dilaton field is given as
\begin{equation}
\label{fundilaton} \frac{8}{3} \partial_z
\left(\frac{L^3e^{3A_s(z)-2\phi} f(z)}{z^3}
\partial_z \phi\right)-
\frac{L^5e^{5A_s(z)-\frac{10}{3}\phi}}{z^5}\partial_\phi V_E=0.
\end{equation}
And the equation of motion for the Maxwell field is
  \bea
\frac{1}{\sqrt{-g^E}}
\partial_\mu \sqrt{-g^E} F^{\mu\nu}=0.\eea It is similar the
equation given by \cite{Andreev:2010bv}. Note that in
(\ref{equation-graviton1})-(\ref{equation-graviton3}) we have only
considered the gravitational configurations with electric charge.

We can see from the equations of motion of the system that once
given a geometric structure $A_s(z)$, one can derive a generic
solution to the system in Einstein frame. The generic solution takes
the form
 \bea
\label{solutionU(1)1}\phi(z)&=&\int_0^z \frac{e^{2 A_s(x)}
\left(\frac{3}{2} \int_0^x y^2 e^{-2 A_s(y)} A_s'(y)^2 \, dy+\phi
_1\right)}{x^2} \, dx+\frac{3 A_s(z)}{2}+\phi _0,\\
A_t(z)&=&A_{t0}+A_{t1} \left(\int_0^z y e^{\frac{2
\phi (y)}{3}-A_s(y)} \, dy\right)\label{solutionU(1)2},\\
f(z)&=&\int _0^zx^3 e^{2 \phi (x)-3 A_s(x)} \left(\frac{A_{t1}{}^2
\left(\int_0^x y e^{\frac{2 \phi
(y)}{3}-A_s(y)} \, dy\right)}{ g_{g}^2 L^2}+f_1\right)dx+f_0\label{solutionU(1)3},\\
V_E(z)&=&\frac{e^{\frac{4 \phi (z)}{3}-2 A_s(z)} }{L^2}\Big(r^2
f''(z)-4 f(z) \left(3 z^2 A_s''(z)-2 z^2 \phi''(z)+z^2 \phi
'(z)^2+3\right)\nonumber\\ &{ }&-\frac{3 z^4 e^{\frac{4 \phi
(z)}{3}-2 A_s(z)} A_t'(z){}^2}{2 L^2
g_{{g}}^2}\Big)\label{solutionU(1)4}, \eea where $\phi_0,\phi_1,
A_{t0}, A_{t1}, f_0, f_1$ are some integration constants, and will
be determined by suitable UV and IR boundary conditions. Generally,
one cannot give the explicit form of $V_E(\phi)$. But for some
special cases, we can use the generating function $A_s(z)$ to obtain
some analytical solutions of the graviton-dilaton-electric field
system. In appendix, we give two analytical solutions which are
generated by this potential reconstruction approach.

 For a consistent check, one can
reproduce the general solution of graviton-dilaton system given by
\cite{Li:2011hp} by turning off the electrical field $A_t(z)$. An
alternative way to check is that one can set $A_s=0, \phi=0, f(z)=1,
A_t(z)=0$ to obtain a constant dilaton potential $V_E(z)=-{12\over
L^2}$ as expected.

\section{Black hole solutions and associated thermodynamics}

\label{blackhole}

 In this section, we will work out general formulas of some
 thermodynamical quantities
for the semi-analytical gravity solutions of the
graviton-dilaton-electrical field system by using
Eq.(\ref{solutionU(1)1})-(\ref{solutionU(1)4}) for given metric
ansatz in Einstein frame (\ref{metric-Einsteinframe}). Here we are
interested in a series of solutions whose UV behavior is asymptotic
$AdS_5$ (aAdS). We also impose the requirements: $f(0)=1$, and
$\phi(z), f(z), A_t(z)$ are regular from $z=0$ to $z_h$. Here $z_h$
is supposed to be the black hole horizon with $f(z_h)=0$. In
addition, we impose  $A_t(z_h)=0$, which is due to the physical
requirement: $A_\mu A^\mu=g^{tt}A_tA_t$ must be finite at the black
hole horizon $z=z_h$.

 The first is how to parameterize the Hawking temperature of the
 black hole solution, which is defined by $\frac{f'(z)}{4\pi}$.
A black hole solution with a regular horizon is characterized by the
existence of a surface $z=z_h$, where $f(z_h)=0$. The Euclidean
version of the solution is defined only for $0 < z < z_h$, in order
to avoid the conical singularity, the periodicity of the Euclidean
time is required as
\begin{equation}
\tau \rightarrow \tau+\frac{4\pi}{|f'(z_h)|}.
\end{equation}
This determines the temperature of the solution as
\begin{equation}\label{temp}
T=\frac{|f'(z_h)|}{4\pi}.
\end{equation}

With help of $f(0)=1$ and black hole horizon $z_h$, there
$f(z_h)=0$, we can fix the integration constants $f_0$ and $f_1$ in
metric function $f(z)$ in eq.(\ref{solutionU(1)3}) as follows.
\bea\label{ffunction} f(z)=1+ \frac{A^2_{t1} }{ g_{g}^2
L^2}\frac{\int_0^z g(x)\left(\int_0^{z_h}g(r)dr \int_r^x
g(y)^{\frac{1}{3}}dy\right)dx}{\int_0^{z_h}g(x)dx} -\frac{\int_0^z
g(x)dx }{\int_0^{z_h}g(x)dx}, \eea where  $f(0)=1$,
$f_1=-\frac{A_{t1}{}^2 }{g_{g_g}^2 L^2}\frac{\int_0^{z_h}dx
g(x)\int_0^xg(y)^{\frac{1}{3}}dy+1}{\int_0^{z_h}g(x)dx}$, and the
function $g(x)$ is defined as \bea \label{xyfunction} g(x)&=&x^3
e^{2 \phi (x)-3 A_s(x)}. \eea One can easily check that $f(z_h)=0$
from Eq.(\ref{ffunction}). One should confirm that there is no other
$z_h$ satisfying  $f(z_h)=0$ in the region $0< z <z_h$. From Eq.
(\ref{temp}), one can easily read out the relation between
temperature and  the black hole horizon from Eq.(\ref{ffunction}) as
\bea\label{Temperature} T=\left|{A^2_{t1}\over 4\pi g_g^2 L^2}\frac{
g(z_h)\int_0^{z_h}g(r)dr\int_r^{z_h}g^{1\over
3}(y)dy-g(z_h)}{\int_0^{z_h}g(x)dx}\right|. \eea

Following the standard Bekenstein-Hawking entropy formula
\cite{entropy-BK}, from the geometry given in
Eq.(\ref{metric-Einsteinframe}), we obtain the black hole entropy
density $s$, which is defined by the area $A_{area}$ of the horizon
\begin{equation}
\label{entrpy} s={\frac{A_{area}}{4 G_5 V_3}=
\frac{L^3}{4G_5}\left(\frac{e^{A_s-\frac{2}{3}\phi}}{z}\right)^3}\Big|_{z_h},
\end{equation}
where $V_3$ is the volume of the black hole spatial directions
spanned by coordinates $x_i$ in (\ref{metric-Einsteinframe}). Note
that the entropy density is determined in terms of horizon area in
Einstein frame.

In order to find out exact expressions about chemical potential and
charge density, we should impose proper boundary conditions on
$A_t$. Expanding $A_t$ near $z=0$, we have
 \bea A_t(z)=
A_{t0}+A_{t1}e^{\frac{2\phi(y)}{3}-A(y)}\left(1+y(\frac{2\phi'(y)}{3}-A'(y))\right)\Big|_{y=0}
z^2 +...
 \label{chemical}\eea
Having considered the boundary condition that $A_t(z_h)=0$, we can
obtain from Eq.(\ref{chemical}) the chemical potential $\mu$ and the
integration constant $A_{t1}$, which is related to the
black hole charge, \bea A_{t0} &=&\mu\\
A_{t1}&=& \frac{\mu}{\int_0^{z_h}y
e^{\frac{2\phi}{3}-A_s(y)}dy}=\frac{\mu}{\int_0^{z_h}
g(y)^{\frac{1}{3}}dy}. \eea Clearly the integration constant
$A_{t1}$ can be determined by chemical potential $\mu$ and horizon
$z_h$. From the coefficient of the $z^2$ term in (\ref{chemical}),
one can obtain the charge density of the black hole configuration.

Before ending this section, we would like to stress that the
integration constants appearing in $\phi(z)$ and $V_E(z)$ do not
occur in the solutions of electrical field $A_t(z)$ and $f(z)$. As
stated in \cite{He:2011hw}, some integration constants will make
contribution to the dilaton potential $V_E(\phi)$, but this does not
affect application of the  potential reconstruction approach to the
EMD system. Based on the builded gravitational configurations, one
 can build up various hQCD models by choosing proper $A_s$. In this paper, we only focus on construction of
holographic model to realize the deconfinement phase transition by
investigating Polyakov loop.  In addition, to be of physical
interest, the builded balck hole solutions should be thermodynamical
and dynamical stable. Therefore it is important to check the
stability of the black hole solutions generally. In  the next
section, we will show that the mass of scalar field in our hQCD
model indeed satisfies the Breitenlohner-Freedman (BF) bound.

\section{The hQCD model with a quadratic correction in warped factor}

\label{sec-solution-hQCD} From various works of constructing
holographic QCD models for describing the heavy quark potential and
the light hadron spectra, we learn that a quadratic background
correction is related to the confinement property, i.e. the linear
quark anti-quark potential \cite{He:2010ye} and the linear Regge
behavior\cite{Dp-Dq}. A positive quadratic correction, $e^{cz^2}$
with $c>0$, in the deformed warp factor of ${\rm AdS}_5$ can help to
realize the linear heavy quark potential \cite{Andreev:2006ct}. A
quadratic dilaton background in the 5D meson action, whose effect in
some sense looks like introducing a negative quadratic correction,
$e^{-cz^2}$, in the warp factor of the ${\rm AdS}_5$ geometry, is
helpful to realize the linear Regge behavior of hadron excitations
\cite{Karch:2006pv}. In our previous study in \cite{Li:2011hp}, it
shows that the hQCD model with a positive quadratic correction in
warped factor is favored. Therefore, we introduce the following hQCD
model with a positive quadratic correction in the  warp factor of
${\rm AdS}_5$ in Eq.(\ref{metric-stringframe}), i.e. we take
\begin{equation} \label{ourmodel}
A_s(z)=  k^2 z^2
\end{equation}
where  $k$ is a parameter related to energy scale of the model,
which will be fixed later. $A_s(z)$ is the generating function of
the solution (\ref{solutionU(1)1})-(\ref{solutionU(1)4}). In terms
of the general expressions for the solution of
graviton-dilaton-electric field system, we can obtain the
configuration of $\phi$ field as
\begin{eqnarray} \label{solu-phi}
\phi(z) &=& \frac{3}{4}  k^2 z^2(1+ H(z)),
\end{eqnarray} where we have set $\phi_0=0$ to keep the metrics (\ref{metric-stringframe}) and
 (\ref{metric-Einsteinframe}) are asymptotical AdS, and $H(z)$ is
\begin{equation} \label{Hc}
H(z)=\, _2F_2\left(1,1;2,\frac{5}{2};2  k^2 z^2\right).
\end{equation}
 With $A_s$ and $\phi$, we can obtain the metric function $f$ as
\begin{eqnarray} \label{solu-f}
f(z)&{}&\equiv  f(z,z_h,\mu)\nonumber\\
&{ }&=1+ \frac{1}{ g_g^2
L^2}\left(\frac{\mu}{\int_0^{z_h}g(y)^{\frac{1}{3}}dy}\right)^2\frac{\int_0^z
g(x)\left(\int_0^{z_h}g(r)dr \int_r^x
g(y)^{\frac{1}{3}}dy\right)dx}{\int_0^{z_h}g(x)dx} \nonumber\\&{
}&-\frac{\int_0^z g(x)dx }{\int_0^{z_h}g(x)dx},
\end{eqnarray}
where
\begin{eqnarray}\label{fc}
g(x)= x^3 e^{\frac{3}{2}  k^2 x^2(1+ H(x))-3k^2 x^2}.
\end{eqnarray}
With the solutions of  $A_s(z)$ and $\phi(z)$, one can obtain the
expression of the electric field \bea\label{A(z)} A_t(z)=\mu +
\frac{\mu}{\int_0^{z_h} g(y)^{\frac{1}{3}}dy} \int_0^z x
e^{\frac{1}{2}k^2 x^2(-1+H(x))}dx. \eea
 Further one can obtain the potential of the dilaton field. The potential includes the contribution from
the electrical field. The form of the potential is too complicated
and therefore we do not present it here. Instead we will discuss the
UV behavior of the dilaton potential to check whether the potential
satisfies the constraint from the 5D Breitenlohner-Freedman (BF)
bound.

The conformal invariance in the UV can be restored when $\phi\sim 0$
at the UV boundary $z \rightarrow 0$. One can expand $\phi(z)$ at UV
boundary $z\sim 0$ as
\begin{equation}
\label{asybehavior} \phi(z\rightarrow 0) \sim \frac{3k^2
z^2}{2}+\frac{3 k^4 z^4}{10}+ \cdots .
\end{equation}
The behavior shown in Eq.(\ref{asybehavior}) is consistent with the
requirement of the asymptotic ${\rm AdS}_5$ near the ultraviolet
boundary.

Through the AdS/CFT dictionary, for any scalar field $\Phi$, we have
\begin{equation}
\label{confdim} \lim_{\Phi\rightarrow 0}V(\Phi) = -\frac{12}{L^2}
+\frac{1}{2L^2} \Delta(\Delta-4) \Phi^2 +O(\Phi^4).
\end{equation}
By using the following relationship
\begin{eqnarray}
\partial^2_{\Phi} V(\Phi)=\frac{\partial r}{\partial \Phi}
\frac{\partial}{\partial r}\left(\frac{\partial r}{\partial \Phi}
\frac{\partial V(z) }{\partial r}\right)= M_{\Phi}^2  +
O(\Phi)+\cdots ,
\end{eqnarray}
one can easily get the conformal dimension of the scalar field. In
Eq. (\ref{confdim}), $\Delta$ is defined as $\Delta( \Delta-4)=
M_{\Phi}^2 L^2$, which is constrained by the BF bound $2< \Delta<
4$. Comparing our case to the standard formula, one should notice
the realtion $\Phi= \sqrt{\frac{8}{3}} \phi$, thus we have
\begin{equation}
M_{\Phi }^2=
 -\frac{4}{L^2}.\end{equation}
 Therefore, the conformal dimension of the dilaton field is
$\Delta=2$ in our model. The dilaton therefore satisfies the BF
bound but does not correspond to any local, gauge invariant operator
in 4D QCD. Although there have been some discussions in recent years
of the possible relevance of a dimension two condensate in the form
of a gluon mass term \cite{Gubarev:2000eu}, it is not clear whether
we can associate $\phi$ with dimension-2 gluon condensate, because
the AdS/CFT correspondence requires that bulk fields should be dual
to gauge-invariant local operators. This situation is the same as
the case in the recent paper \cite{Li:2011hp}. Note that in the
potential reconstruction approach, the electrical field does not
change the configuration of dilaton and the UV behavior of the
dilaton potential.

\begin{figure}[h]
\begin{center}
\epsfxsize=7.5 cm \epsfysize=6.5 cm \epsfbox{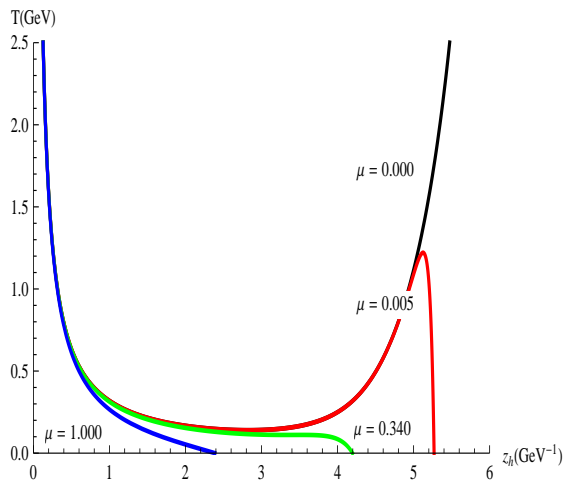}
\end{center}
\caption{The black hole temperature versus horizon $z_h$ with
different chemical potential. When $\mu=0$, the temperature behavior
is similar to the case considered in\cite{Li:2011hp}. When $\mu
>\mu_c$ the temperature  monotonically decreases to zero as $z_h$
increases. While $0 <\mu <\mu_c$, the temperature decreases to a
minimum at $z_m$ and then grows up to a maximum at $z_M$ and then
decreases to zero monotonically. When $\mu=\mu_c$, one has
$z_m=z_M$. $z_m$ and $z_M$ are marked explicitly in figure 2. In our
model, we fix $ k=0.3 \text{GeV}$ and accordingly the critical
chemical potential is $\mu_c=0.340 \text{GeV}$. }
 \label{T-zh}
\end{figure}

\begin{figure}[h]
\begin{center}
\epsfxsize=6.5 cm \epsfysize=5.5 cm \epsfbox{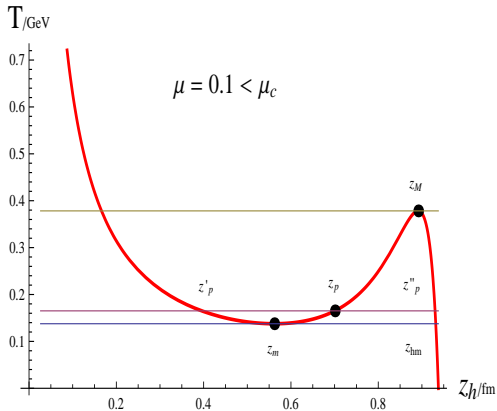} \
\end{center}
\caption{The temperature of the black hole with $\mu=0.1\text{GeV}$.
The three black hole solutions with horizon $z'_p$, $z_p$ and
$z''_p$ have the same temperature. The black hole with $z_m
<z_p<z_M$ is thermodynamically unstable. In the figure we take $g_g
L=1, k=0.3 \text{GeV}$.} \label{Small-mu-example}
\end{figure}

Once the metric function $f$ is given, one can easily find the
relation between the black hole temperature and the horizon radius
$z_h$. In figure 1 we plot the relation between the temperature and
horizon with different chemical potential. One can see from figure
\ref{T-zh} that the black hole temperature can not reach zero
temperature if the chemical potential is zero. On the other hand, if
one turns on a nonvanishing chemical potential, the lowest
temperature can reach zero, this case corresponds to an extremal
black hole. In the case with a nonvanishing chemical potential, as
one can see from the figure that the behavior of the temperature
heavily depends on the value of chemical potential. Basically the
behavior of the temperature can be classified into two cases: when
chemical potential $\mu > \mu_c$, the temperature decreases
monotonically to zero as $z_h$ increases, while if $\mu <\mu_c$, it
decreases monotonically to a minimum at $z_m$, then goes to a
maximum at $z_M$ and then decreases to zero. The critical chemical
potential depends on the parameter $k$. In our model, we take
parameter $k=0.3 \text{GeV}$, which is consistent with the lattice
data for equation of state, heavy quark potential and so
on~\cite{Li:2011hp}. We have checked that if one changes  the value
of $k$, our conclusions in this paper do not change qualitatively.
In the case of $k=0.3{\rm GeV}$, the corresponding critical chemical
potential $\mu_c=0.34 {\rm GeV}$. Here the critical chemical
potential is determined by
$T({\mu=\mu_c,z_h=z_m})=T({\mu=\mu_c,z_h=z_M})$. And $z_m$ and $z_M$
have been presented in figure \ref{Small-mu-example}.

From the behavior of the temperature, one can see that the black
hole solution is locally thermodynamical stable when $\mu>\mu_c$,
because in this case, the heat capacity of the solution is positive
(Note that the horizon size of the black hole is $1/z_h$). On the
other hand, the case with $\mu <\mu_c$ has to be discussed
separately. To clearly see the stability of the black hole solution
in the case $\mu <\mu_c$, in figure \ref{Small-mu-example} we plot
the temperature versus the horizon $z_h$ in the case $\mu=0.1 {\rm
GeV} <\mu_c$. We can see from the figure that the background black
hole is thermodynamical unstable in the region $z_m<z_h <z_{M}$,
where $z_m$ and $z_M$ are the black hole horizons, respectively,
corresponding to the minimal and maximal temperatures. In this
region, the heat capacity  of the black hole is negative. The black
hole solutions in the regions $z_h <z_m$ and $z_h>z_M$ are
thermodynamical stable. When $\mu=\mu_c$, $z_m$ and $z_M$ are
degenerated to one point.

The existence of the critical chemical potential plays a crucial
role to realize the critical point in QCD phase diagram. We will see
this in the next section by calculating the heavy quark potential in
this hQCD model.

\section{Heavy quark potential, Polyakov loop and QCD Phase diagram}

To investigate some properties of the hQCD model constructed in the
previous section, let us study an infinitely heavy quark-antiquark
pair, at distance $r$ from each other. It is interesting to
investigate how the free energy of such a system changes with
 temperature and chemical potential by using the holographic
description of loop
operators~\cite{Maldacena:1998im,Rey:1998bq,Polyakov:1997tj}. The
free energy is a proper quantity to describe the
confinement/deconfinement phase transition.

\subsection{Heavy quark potential and Polyakov loop}

\begin{figure}[h]
\begin{center}
\epsfxsize=5.5 cm \epsfysize=5.5 cm \epsfbox{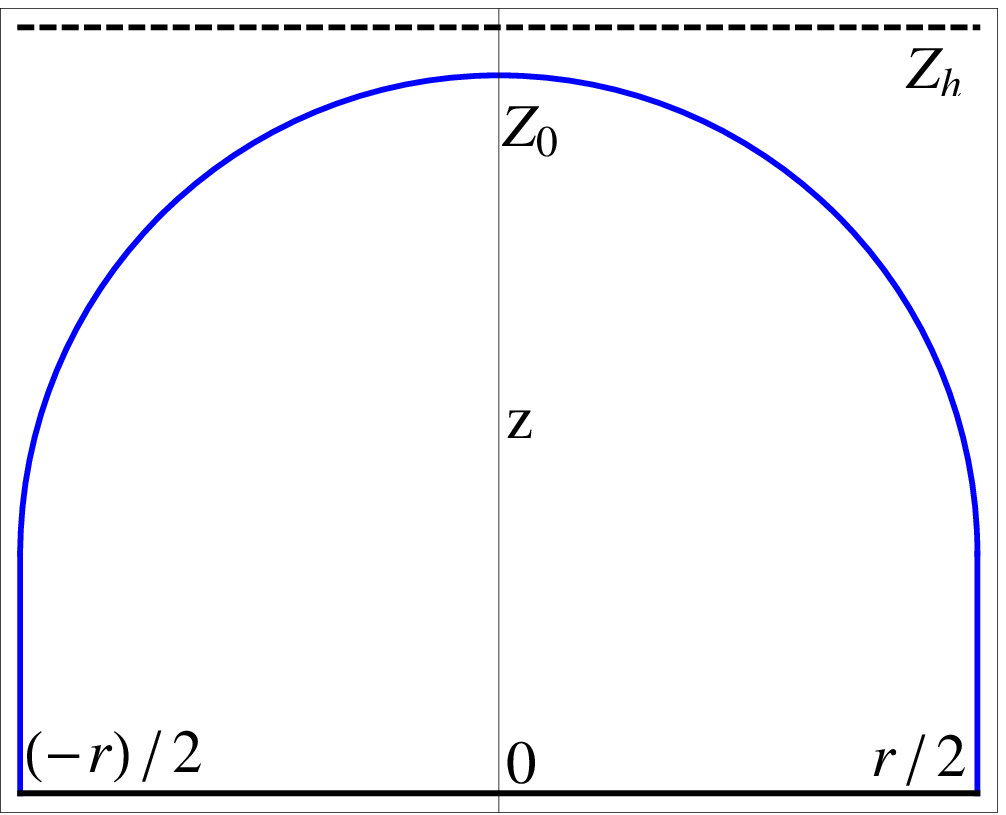} \hspace*{0.1cm}
\epsfxsize=5.5 cm \epsfysize=5.5 cm \epsfbox{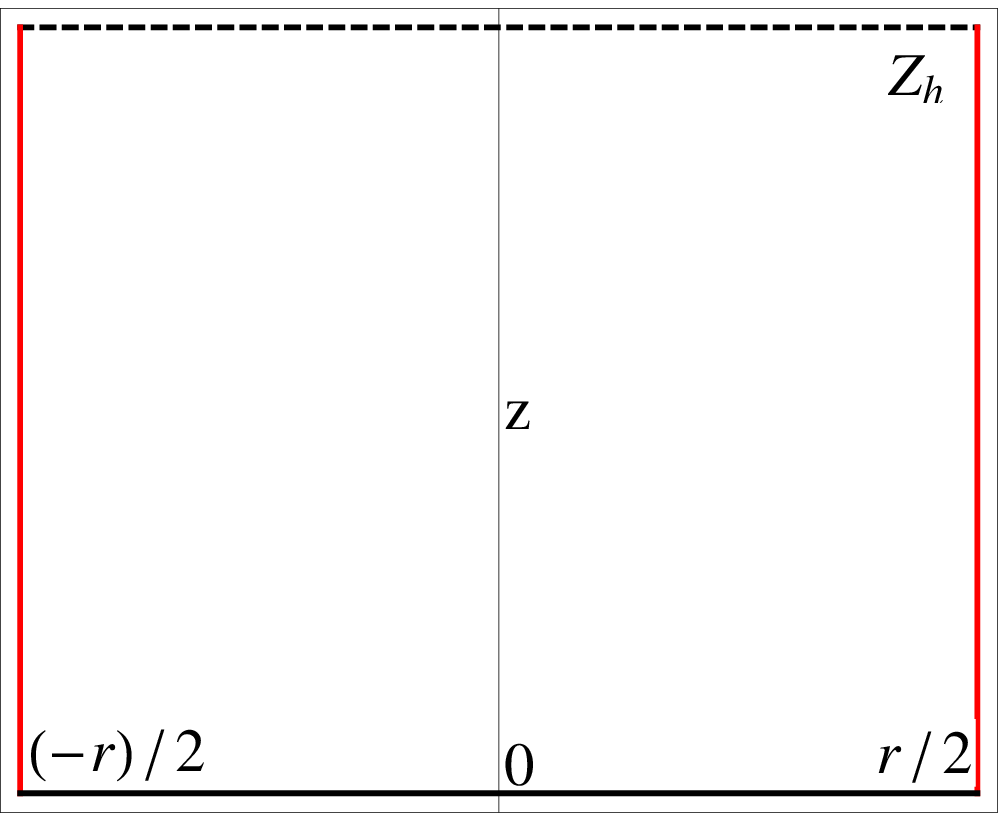} \vskip -0.05cm
\hskip 0.15 cm
\textbf{( a ) } \hskip 6.5 cm \textbf{( b )} \\
\end{center}
\caption{The two configurations represent the confinement phase and
deconfinement phase. In Figure (a)  there is a fundmental string
connecting with two static quarks on the boundary. This
configuration denotes the confining phase. In Figure (b) one string
beaks into two segments, both of them  extend to the horizon of the
black hole. This configuration corresponds to the deconfinement
phase.} \label{configration}
\end{figure}

 At finite
temperature, the free energy $F(r,T)$ of an infinitely heavy
quark-antiquark pair at distance $r$ can be obtained in QCD from the
correlation function of two Polyakov loops
\cite{Polyakov:1997tj}
\begin{equation}\label{twopolyakov}
\langle{ \mathcal{P}}(\vec x_1) { \mathcal{P}}^\dagger(\vec x_2)
\rangle= e^{-\frac{1}{T} F(r,T)+\gamma(T)}
\end{equation}
with $r=|\vec x_1-\vec x_2|$ and $\gamma(T)$ a normalization
constant. Moreover, the vacuum expectation value of a single
Polyakov loop
\begin{equation}\label{polyakov}
\langle{ \mathcal{P}} \rangle= e^{- \frac{1}{2 T}  F^\infty(T)}
\end{equation}
($ F^\infty(T)=F(r=\infty,T)$ and the normalization factor is
neglected) is an order parameter for the deconfinement transition of
a gauge theory~\cite{Polyakov:1997tj}. Within the gauge/string
duality approach, we can attempt a calculation of the expectation
values in \eqref{twopolyakov} and \eqref{polyakov} by considering a
fundamental string configuration with Polyakov loop on the boundary.
 The Nambu-Goto action of the string is
\begin{equation}\label{NG}
S_{\mbox{\tiny NG}}=\frac{1}{2\pi \alpha^\prime} \int d^2\xi
\sqrt{\det\left[g_{ab}\right]}=\frac{1}{2\pi \alpha^\prime} \int
d^2\xi \sqrt{\det\left[g_{MN}\left(\partial_a
X^M\right)\left(\partial_b X^N\right)\right]}
\end{equation}
In terms of the geometry background $g_{MN}$ given by
(\ref{metric-stringframe}), the induced metric $g_{ab}$ can be
expressed as \bea\label{induceM} g_{ab}=\left(
\begin{array}{cc}
 -\frac{e^{2 A_s(z)} f(z)}{z^2} & 0 \\
 0 & \frac{e^{2 A_s(z)} \left(\frac{1}{f(z)}\frac{{d z}^2}{  d x^2}+1\right)}{z^2}
\end{array}
\right) ,\eea where we have chosen the static gauge $\xi^0=t$ and
$\xi^1=x$. We are interested in the configurations of two static
quarks on the boundary. The configurations should obey the boundary
conditions: $z(x=0)=z_0$ and $z'(x=0)=0$ and
$z(x=\pm\frac{r}{2})=0$. Here the prime stands for the derivative
with respect to $x$.
  There are two independent configurations as shown in figure
\ref{configration} which satisfy the boundary conditions. Figure
\ref{configration}(a) stands for the case with two static quarks
linked by a fundamental string. This configuration describes the
confining phase. Figure \ref{configration}(b) describes the case
where the string is broken into two segments, each of them extends
to the horizon of black hole.  This configuration stands for two
static decoupled quarks, which corresponds to the deconfinement
phase.  The configuration in figure (a) corresponds to the
``Minkowski embedding", while figure (b) to the ``black hole
embedding" in the $D_3/D_7$ setup~\cite{arXiv:1101.0618}.

From \eqref{NG} and \eqref{induceM}, we obtain  the free energy of
the string configuration
\begin{equation}\label{Ftemp}
F(r,T)=\frac{g_p}{\pi}\int_{-r/2}^0 dx
\frac{e^{k^2z^2}}{z^2}\,\sqrt{f(z)+\left(z^\prime\right)^2},
\end{equation}
where $g_p=\frac{L^2}{\alpha^\prime}$.  One can see from
\eqref{Ftemp} that the configuration of the string satisfies the
constraint
\begin{equation}
{
H}=\frac{e^{k^2z^2}}{z^2}\,\frac{f(z)}{\sqrt{f(z)+\left(z^\prime\right)^2}},
\end{equation}
where  $H$ is a conservation quantity, the Hamiltonian of the string
configuration.

 For later convenience, we  express $F(r,T)$ in terms of
$z_0$ and $f_0=f(z)|_{z=z_0}$. Defining $v={z\over z_0}$, and
subtracting the UV ($v\to 0$) divergence in (\ref{Ftemp}), which
corresponds to subtract the infinite quark and antiquark mass in
four dimensional QCD~\cite{Maldacena:1998im}, we obtain the
regularized free energy for the string configuration in the
confining phase

\begin{equation}\label{F}
\hat{F}(\lambda,
T)=\frac{g_p}{\pi\lambda}\left[-1+\int_0^1\frac{dv}{v^2}\left(\frac{e^{\lambda^2
v^2}}{\tau(v)}-1\right)\right],
\end{equation}
where $\hat F=F/k$,  $\lambda=k\,z_0$ and
\begin{eqnarray}\label{divergent}
\tau(v) & = & \sqrt{1-\frac{f_0}{f(z_0
v)}\,v^4e^{2\lambda^2\left(1-v^2\right)}}\,\,.
\end{eqnarray}
The distance $\hat r= k r$ can also be parameterized by $z_0$,
\begin{equation}\label{r}
\hat r(z_0)=2 \lambda^2\sqrt{f_0} \int_0^1 dv\, \frac{v^2\,
e^{\lambda^2(1-v^2)}}{\tau(v)f(z_0 v)}.
\end{equation}

Next we calculate the regularized free energy for the configuration
in the decconfinement phase. In this case, one has $z_0=z_h$. Using
the gauge $\xi^0=t$ and $\xi^1=z$ and $x'(z)=0$, we obtain the
regularized free energy
\begin{equation}\label{Finf}
\hat F^\infty (\hat{T})=
\frac{g_p}{\pi}\left[-\frac{1}{\hat z_h}+\int_0^{\hat
z_h}\frac{d\hat z}{\hat z^2}\left(e^{\hat
z^2}-1\right)\right]+\zeta(\hat\mu,\hat T),
\end{equation}
where $\hat{T}=T/k$ and $\hat z=kz$. $\zeta(\hat\mu,\hat T)$ is a
term related to the regularization procedure, and we identify the
maximum of the free energy $\hat F(r,T)$ with $\hat F^\infty$ to fix
the regularization term following~\cite{Colangelo:2010pe}.

\subsection{QCD phase diagram}
In this subsection, we will discuss the phase diagram of the hQCD
model by analyzing the regularized free energies obtained in the
previous subsection for those two different configurations.


To analyze the behavior of $\hat{r}(z_0)$ is helpful to distinguish
two different phases.  Note that there might exist a divergence in
(\ref{r}) for some parameters due to a vanishing $\tau(v)$. An
infinite $\hat{r}$ in figure 2(a) implies that two static quark are
still coupled even with infinite separation. This means that the
case with the infinite $\hat{r}$ is in the confining phase. If there
does not exist any divergence in $\hat{r}(z_0)$, one can find a
maximum $\hat{r}_{\rm max}$  at some $z_0$ with fixed chemical
potential $\mu$ and temperature $T$. Beyond $\hat{r}_{\rm max}$, the
string configuration in the confining phase becomes unstable, and
the deconfinement phase transition happens in this case.

\begin{figure}[h]
\begin{center}
\epsfxsize=7.5 cm \epsfysize=6.5 cm \epsfbox{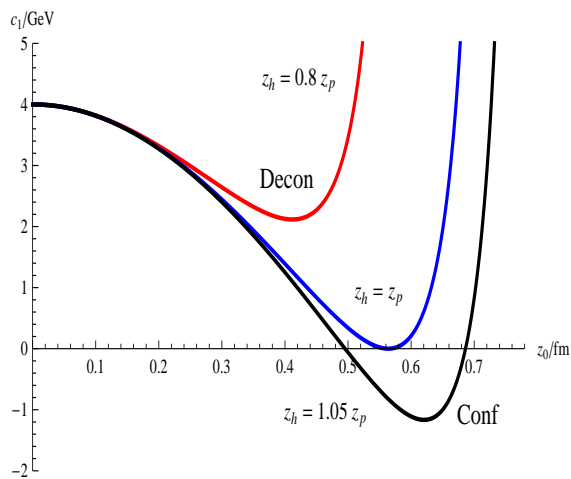}
\\
\end{center}
\caption{This plot shows the behavior of $c_1$ with respect to $z_0$
with three different black hole horizons $z_h$ and a fixed chemical
potential $\mu=0.1\text{GeV}$  The solid red curve represents the
deconfinement phase, the solid black curve corresponds to the
confinement phase, and the solid blue one denotes the deconfinement
phase transition. $z_p$ is the critical horizon which leads $c_1$ to
have a single zero root.
 $z_h<z_p$ and $z_h\geq z_p$ correspond to the deconfinement
phase and confinement phase respectively. Here we take $g_g L=1$ and
$ k=0.3 \text{GeV}$.} \label{c1}
\end{figure}

We can see from (\ref{divergent}) that when $v=1$, one has $\tau=0$.
This means that the divergence in (\ref{r}) appears when $\tau=0$.
 Note that other factors in (\ref{r}) do not lead to any divergence.
 Let us first analyze the behavior of $\tau(v)$ near $v=1$.
Expanding the quantity in the square root in $\tau(v)$ at $v=1$, one
has
 \bea
\tau(v)\sim \sqrt{c_1 (1-v)+c_2(1-v)^2+{\cal O}((1-v)^2)}, \eea
where $c_1$ and $c_2$ are two expansion coefficients. Note that the
following relations
\begin{eqnarray}\label{CC1}
\int_0^1dv \frac{1}{\sqrt{1-v}}&=&2,\nonumber\\
 \int_0^1dv
\frac{1}{1-v}&=&\infty.
\end{eqnarray}
We can see that if $c_1=0$, the integration in (\ref{r})) is
divergent, otherwise it always gives a finite result. Furthermore we
find that $c_1$ is of the following form
\begin{equation}\label{Polec1}
c_1\sim\frac{z_0}{f(z_0,z_h,\mu)} \frac{df(z_0,z_h,\mu)}{dz_0}+8 k^2
z_0^2-4,
\end{equation} with positive $z_h,\mu$ and $0<z_0<z_h$.
Note that the first term in the right hand side in (\ref{Polec1}) is
negative due to $\frac{\partial f(z,z_h,\mu)}{\partial z}<0$ in the
region $z_0<z_h$, and the second term is always positive. Therefore,
in the case with fixed $\mu$, if $z_h$ is large enough, $c_1$ can
reach zero at some $z_0 (<z_h)$. At that point, $\hat{r}(z_0)$ is
divergent. On the other hand, in some region of parameters $z_h$ and
$\mu$, $c_1$ is always positive definite and $\hat{r}(z_0)$ is then
 finite in that region. Let us stress again that the case with
a vanishing $c_1$ is always in the confining phase, while  the case
with positive definite $c_1$ corresponds to deconfinement phase. In
figure \ref{c1} we show the behavior of $c_1$ with respect
 to $z_0$ for various $z_h$ and a fixed chemical potential $\mu=0.1\text{GeV}$.

\begin{figure}[h]
\begin{center}
\epsfxsize=7.5 cm \epsfysize=6.5 cm \epsfbox{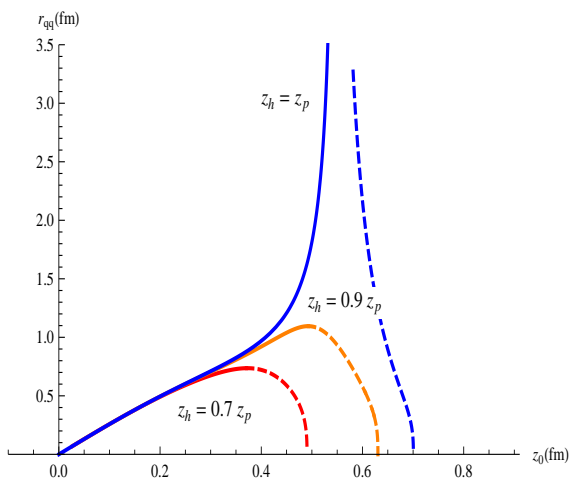}
\\
\end{center}
\caption{This plot shows the relation between the distance between
two quarks and $z_0$ with a fixed $\mu=0.1\text{GeV} $. The blue
curve corresponds to the configuration given in figure
\ref{configration} (a) which stands for confining phase, while the
red  and pink curves correspond to the configurations given figure
\ref{configration} (b). When $z_h=z_p$, $r(z_0)$ can be infinity at
some $z_0$ in $0<z_0<z_h$. The dashed curves are related to
configurations in unstable region.} \label{RZ}
\end{figure}

One can also obtain free energy $F$ of the static quark-antiquark
potential in two different phases through (\ref{F}). In figure
\ref{RZ}, we plot the function $r(z_0)$ with a fixed $\mu=0.1 {\rm
GeV}$. We see clearly that when $z_h< z_{p}$ and then $c_1> 0$,
$r(z_0)$ has a maximum $r_{\rm max}$. This case corresponds to the
deconfinement phase.  On the other hand, the $z_h\geq z_{p}$ cases
correspond to the confining configuration in which  there exists a
 divergence in the blue curve $r(z_0)$.
 In the confining phase, if the separation $r(z_0)$ of two static quarks can go to
infinity, the phase is called permanent confinement phase. The red
and pink solid curves correspond to the deconfinement case. In these
cases, there exist some maximums in the curves $r(z_0)$ in
$0<z_0<z_h$, beyond those maximums there are no stable
configurations of two coupled quarks.

\begin{figure}[h]
\begin{center}
\epsfxsize=7.5 cm \epsfysize=6.5 cm \epsfbox{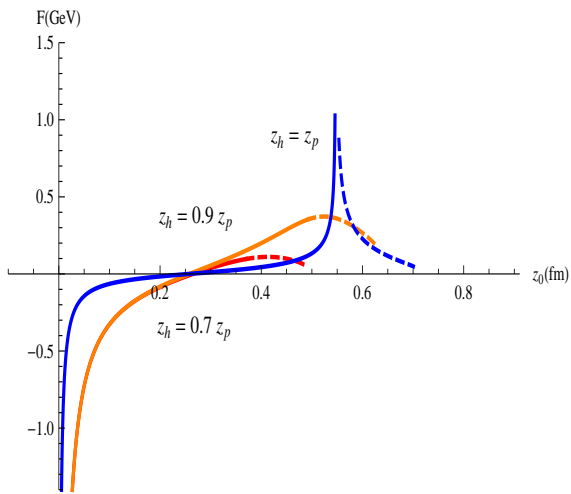} \\
\end{center}
\caption{The free energy $F$ of string configurations versus $z_0$
with a fixed $\mu=0.1\text{GeV}$.  The solid blue curve corresponds
to the confinement phase and the other two to the deconfinement
phase. Here $g_g L=1,g_p=1, k=0.3 \text{GeV}$. The dashed curves are
related to configurations in unstable region.} \label{VQQ-z}
\end{figure}

\begin{figure}[h]
\begin{center}
\epsfxsize=7.5 cm \epsfysize=6.5 cm \epsfbox{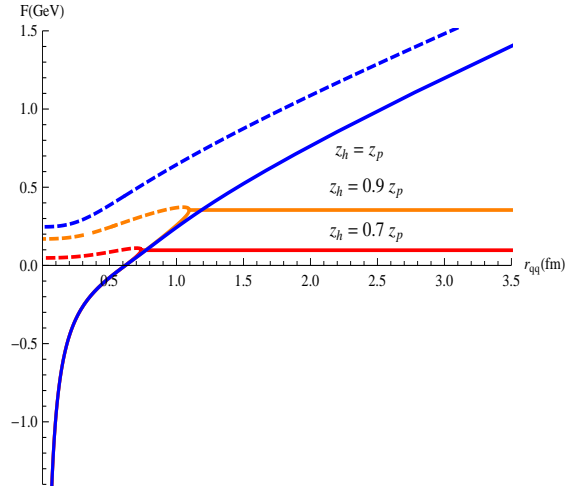} \\
\end{center}
\caption{This plot show the  free energy $F(r,T)$  versus the
distance $r$ between two quarks with a fixed $\mu=0.1$.  The solid
blue curve corresponds to the confinement phase and the other two
 to the deconfinement phase.
In the confining phase the free energy $F$ goes to infinity when $r
\to\infty$. In the deconfining phase there is a maximum in
$0<z_0<z_h$. Here $g_g L=1, g_p=1, k=0.3 \text{GeV}$. The dashed
curves are related to configurations in unstable region.}
\label{VQQR}
\end{figure}

In figure \ref{VQQ-z}, we plot the free energy with respect to $z_0$
with a fixed $\mu=0.1 {\rm GeV}$. We can see from the figure that in
the deconfinement case, the free energy has a maximum in
$0<z_0<z_h$. We identify the maximum of $F(r,T)$ with
$F^{\infty}(T)$ following the strategy given in
\cite{Colangelo:2010pe}.  In figure \ref{VQQR} the free energy is
plotted with respect to the distance between two quarks.  We see
from the figure that the potential with $z_h=z_p$ can go to infinity
when $r\to \infty$.  This means the vacuum expectation value (VEV)
of polyakov loop (\ref{polyakov}) in this phase is vanishing, which
implies the configuration is in the confining phase.  On the other
hand, the free energy $F$ in the cases $z_h=0.7z_p$ and $z_h= 0.9
z_p$ is finite when $r\rightarrow \infty$. This implies that in this
case the
 VEV of Polyakov loop (\ref{polyakov}) is not vanishing and the
 configuration is in the deconfinement phase.

Through the above analysis, we see that the key point of finding
phase transition in the hQCD model is to find the critical horizon
$z_h=z_p$, for which $\hat r(z_0)$ is divergent. The strategy is the
same as the one employed in works
\cite{Colangelo:2010pe,hep-ph/0611304}. The numerical results in
figure \ref{RZ} show that one can fix the critical horizon $z_p$ and
then the phase transition temperature with a fixed chemical
potential $\mu$ through (\ref{Polec1}).

Let us first discuss the case with $\mu <\mu_c$. In this case the
phase transition is a first order one. To confirm this, let us see
the phase transition from the Polyakov loop perspective
\cite{arXiv:0812.0792}.
 From our numerical study, one can see that
 the critical horizon $z_p$ determined by free energy of heavy quark pair
 always lives in $z_m<z_p<z_M$ when $\mu <\mu_c$. The phase transition temperature
 corresponds to the black hole having the horizon $z_p$. But the black hole with $z_p$ is thermodynamical
 unstable.  Furthermore one can see from
 figure \ref{Small-mu-example} that there exist another two black hole
 solutions with horizon $z'_p$ and $z''_p$, which have  the same temperature
 as the black hole with horizon $z_p$. The two black hole solutions
 with $z'_p$ and $z''_p$ are thermodynamical stable. They correspond
 to two different phases, deconfinement phase and confinement phase,
 respectively. The deconfinement phase transition happens from the black hole
 with $z''_p$ to the one with $z'_p$ at the transition temperature.

\begin{figure}[h]
\begin{center}
\epsfxsize=6.5 cm \epsfysize=5.5 cm \epsfbox{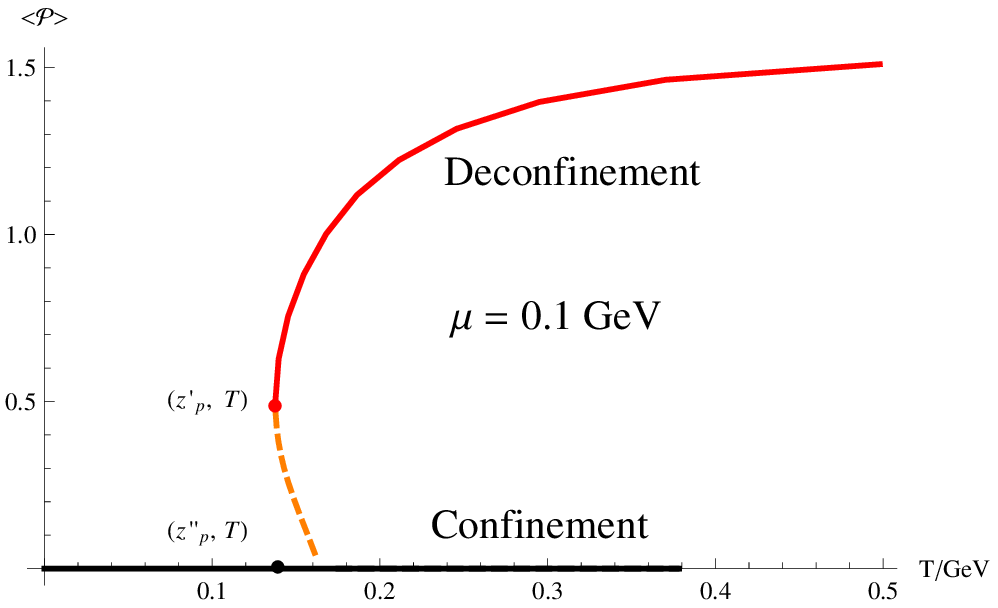}
\hspace*{0.1cm} \epsfxsize=6.5 cm \epsfysize=5.5 cm
\epsfbox{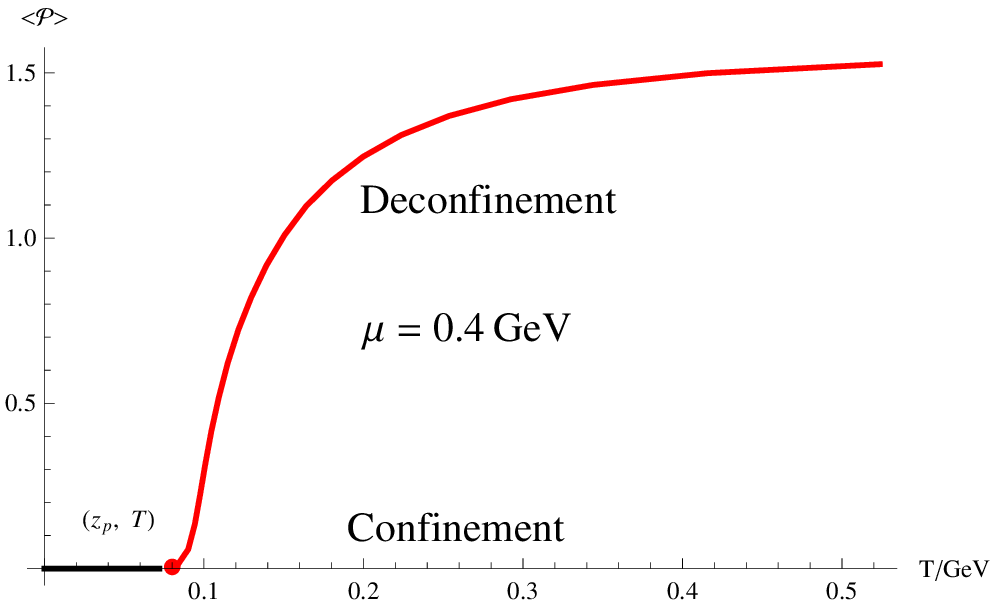} \vskip -0.05cm \hskip 0.15 cm
\textbf{( a ) } \hskip 6.5 cm \textbf{( b )} \\
\end{center}
\caption{The vacuum expectation of a single Polyakov loop versus
temperature. The left panel shows the case with $\mu <\mu_c$ and the
right one with $\mu >\mu_c$.  In (a) the dashed pink curve shows the
behavior of $ \langle \mathcal {P}\rangle $ in the unstable region
$z_m<z_h<z_M$. Here $g_p=1, k=0.3 \text{GeV}$.} \label{Fqq}
\end{figure}

After fixing the phase transition temperature, next we calculate the
VEV of a single Polyakov loop by using (\ref{Finf}), which is a
function of temperature. We find that when $z_h >z_p$, $\langle
{\cal P}\rangle$ is always vanishing, while $\langle {\cal
P}\rangle$ does not as $z_h<z_p$.  Therefore there must be a jump of
$\langle \mathcal {P}\rangle $ from $0$ to a nonvanishing one in
$\langle \mathcal {P}\rangle -T$ plane as shown in figure
\ref{Fqq}(a). The black dot and red dot in the figure represent the
black hole solutions with $z''_p$ and $z'_p$, respectively. We can
clearly see from the figure that the phase transition is a first
order one in this case.

Now we discuss the case with  $\mu \geq \mu_c$. In this case, the
temperature is a monotonic function of horizon $z_h$ as shown in
figure \ref{T-zh}. After fixing the phase transition temperature
through (\ref{Polec1}), we calculate the VEV of the a single
Polyakov loop by using (\ref{Finf}). We find that when $z_h >z_p$,
$\langle {\cal P}\rangle=0$, which corresponds in the confining
phase, while $\langle {\cal P}\rangle$ increases monotonically from
$0$ as $z_h \le z_p$. The behavior of $\langle {\cal P}\rangle$ is
shown in \ref{Fqq}(b). This case can be interpreted as a continuous
transition~\cite{arXiv:1111.4953}. The continuity of $\langle {\cal
P}\rangle$ at the phase transition point implies that the
deconfinement transition might be a crossover in the heavy quark
limit~\cite{Fodor:2009ax}.

\begin{figure}[h]
\begin{center}
\epsfxsize=7.5 cm \epsfysize=5.5 cm \epsfbox{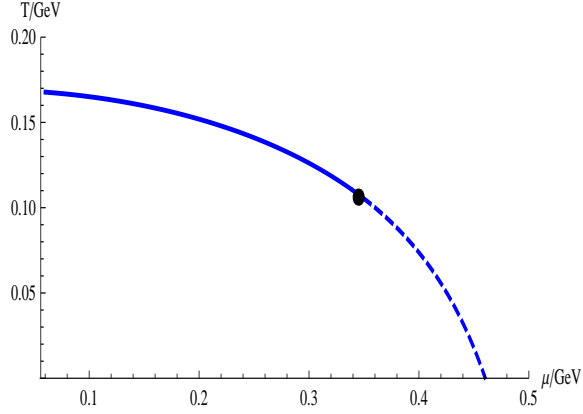}
\end{center}
\caption{The phase diagram of the hQCD model constructed in this
paper. The dashed blue curve stands for the continuous phase
transition and the solid blue one for the first order phase
transition. The black dot represents the critical point. Here
$g_p=1, k=0.3 \text{GeV}$.}\label{PhaseDia1}
\end{figure}

Combing the above analysis, we plot a naive $T-\mu$ phase diagram in
figure \ref{PhaseDia1} for our hQCD model.  In the figure the black
dot denotes the critical point with $\mu = \mu_c$. When $\mu
<\mu_c$, the deconfinement transition is a first order one, while it
is a continuous transition as $\mu >\mu_c$. This phase diagram
 agrees with the expectation from effective field theory
\cite{hep-ph/0402234,nucl-th/0403039} and recent lattice QCD
simulation \cite{arXiv:1111.4953}. By using effective theory, the
recent lattice QCD simulation \cite{arXiv:1111.4953} studies the
deconfinement transition of QCD with heavy quark. It is found that
the deconfinement transition is first order in small chemical
potential region, while it is an analytic crossover in large $\mu$
region.  Our results therefore are consistent with the lattice
simulation \cite{arXiv:1111.4953}. Furthermore, note that the work
\cite{Colangelo:2010pe} considered a hQCD model by introducing a
warped factor to deform the AdS Reinsser-Nordstr\"om black hole
background and concluded that the deconfinement phase transition is
a continuous transition. Differing from  the work
\cite{Colangelo:2010pe}, we have constructed a self-consistent
gravitational configuration by considering the back reaction of
dilaton field and a critical point has been found in the phase
diagram.

\section{Conclusion and discussion}
The gravity/gauge theory duality is a powerful tool to investigate
strongly coupling systems. In this spirit, one of dreams is to build
a holographic description dual to a real low energy QCD theory. In
this paper, by using the potential reconstruction approach
\cite{Li:2011hp,He:2011hw}, we have given a generic formulism to
find a series of asymptotically AdS black hole solutions for the
Einstein-Maxwell-dilaton system. In this approach, the back reaction
of dialton and Maxwell field is taken into account. In this sense,
our approach avoids some shortcomings in the literature in
constructing holographic QCD models.

Based on the approach, we have constructed a self-consistent
gravitational configuration to describe some properties of low
energy QCD theory.  The gravitational configuration includes a
quadratic term in warped factor of bulk metric. The behavior of
temperature of the black hole configuration is similar to that of an
AdS  Reissner-Nordstr\"om black hole. The quadratic term in warped
factor plays an important role in the hQCD model. By calculating
heavy quark potential and Polyakov loop in this hQCD model, we have
analyzed the phase structure of the model. It has been found that
there exists a critical point in $T-\mu$ phase diagram. When the
chemical potential $\mu <\mu_c$, the deconfinement phase transition
is a first order one, while it is a continuous transition when
$\mu>\mu_c$.  This phase diagram is agreement with results from
effective field theory \cite{hep-ph/0402234,nucl-th/0403039} and
recent lattice QCD simulation \cite{arXiv:1111.4953}. In our model,
the value of the critical chemical potential depends on the model
parameter $k$. In our discussions, $k=0.3 {\rm GeV}$ and then
$\mu_c=0.34 {\rm GeV}$. We have checked other values of $k$ and our
results do not change qualitatively and the conclusions still holds.

In this work we have only studied the deconfinement transition of
the hQCD model in heavy quark limit, by studying Polyakov loop in
the black hole background presented in this paper. It would be of
great interest to further investigate other aspects of the hQCD
model. For example, it is required in the model to further study the
spectra of hadrons, chiral phase transition
\cite{Evans:2011mu,Evans:2011tk,Evans:2011eu}, hydrodynamical
properties of QGP, and
 color flavor locked phase \cite{arXiv:0909.1296,arXiv:1101.4042,hep-ph/0302142}, etc.. In addition, it would be also
interesting to construct thermal gas solution in the EMD system and
to discuss the Hawking-Page phase transition between the black hole
solution and the thermal gas solution.

\vskip 1cm \noindent {\bf Acknowledgments}: The authors thank T.
Hatsuda, Mei Huang, Hong Mao, Defu Hou, Tamal Mukherjee, Dawei Pang,
Fukun Xu, Jun Tao, Shunjin Wang, Junbao Wu, Haitang Yang, Qishu Yan
and Yi Yang for valuable discussions. The authors specially thank
Elias Kiritsis for his helpful comments and suggestions, which help
us to improve this paper. S.H. appreciates the hospitality of
institute of high energy physics, CAS and Physics Department of
Sichuan University at various stages of this work. This work is
supported in part by grants from NSFC (No. 10821504, No. 10975168
and No. 11035008). This work is also supported partially by FRFCU,
No. 2011RC22.

\appendix
\renewcommand{\theequation}{\thesection.\arabic{equation}}
\addcontentsline{toc}{section}{Appendices}
\section*{APPENDIX}
\section{Two analytical black hole solutions in EMD system}

In this appendix, we give two analytical black hole solutions of EMD
system by using Eqs.(\ref{solutionU(1)1})-(\ref{solutionU(1)4}).
Here we are interested in the solutions whose UV behavior is
asymptotic $AdS_5$. We  impose the constraint $f(0)=1$ at $z= 0$,
and require $\phi(z)$ and  $f(z)$ to be regular in the region $z \in
[0, z_h]$, where $z=0$ is the AdS boundary and $z_h$ corresponds to
the horizon of black hole solution.

We give black hole solutions in Einstein frame. The metric in
Einstein frame takes the following form
\begin{eqnarray} \label{Ametric-Einsteinframe}
ds_E^2&=&\frac{L^2 e^{2A_s-\frac{4\phi}{3}}}{z^2}\left(-f(z)dt^2
+\frac{dz^2}{f(z)}+dx^{i}dx^{i}\right)\\
&=&\frac{L^2 e^{2A_E}}{z^2}\left(-f(z)dt^2
+\frac{dz^2}{f(z)}+dx^{i}dx^{i}\right),
\end{eqnarray}
where $A_E(z)=A_s(z)-\frac{2\phi(z)}{3}$.

The first set of exact solutions is
\begin{eqnarray}
A_E(z)&=&\log \left(\frac{z }{z_0\sinh(\frac{z}{z_0})}\right),\nonumber\\
f(z)&=&1-\frac{4 V_{11}}{3}(3\sinh(\frac{z}{z_0})^4+2\sinh(\frac{z}{z_0})^6)+\frac{1}{8} V_{12}^2 \sinh\left(\frac{z}{z_0}\right)^4,\nonumber\\
\phi(z)&=&\frac{3 z}{2 z_0},\nonumber\\
A_t(z)&=& \mu -\frac{2g_g L}{z_0} V_{12}
\sinh\left(\frac{z}{2z_0}\right)^2,
\end{eqnarray}
where $z_0$ and $\mu$ are two  integration constants, $V_{11}$ and
$V_{12}$ are two constants from the dilaton potential,  and $g_g$ is
gauge coupling. The dilaton potential is given by
\bea\label{dilatonpotential1}
V_E(\phi)&=&-\frac{12+9\sinh^2\left(\frac{2\phi}{3}\right)
+16V_{11}\sinh^6\left(\frac{\phi}{3}\right)}{L^2}+\frac{V_{12}^2
\sinh^6\left(\frac{2 \phi }{3}\right)}{8 L^2}. \eea
 If turn off the gauge field, one can
reproduce one solution of the graviton-dilton system given in
\cite{Li:2011hp}\cite{He:2011hw}. On the other hand, if set
$V_{11}=0$ and $ V_{12}=0$,  one can reach the 5D BPS solution in
\cite{He:2011hw}.

 The second set of exact solutions is \bea
A_E(z)&=&-\log \left(1+\frac{z}{z_0}\right),\\
f(z)&=& 1-V_{21} \left(\frac{z^7}{7 z_0^7}+\frac{z^6}{2 z_0^6}+
+\frac{3 z^5}{5 z_0^5}+\frac{z^4}{4 z_0^4}\right)\nonumber\\
&&+\frac{\rho^2 z_0^8}{ g_g^2 L^2}\Big( \frac{5 z^8 }{32
z_0^8}+\frac{z^{10} }{60
 z_0^{10}} +\frac{z^9}{12 z_0^9}+\frac{11 z^7 }{84 z_0^7}
+\frac{z^6 }{24z_0^6}\Big),\\
\phi(z)&=&3 \sqrt{2} \sinh
^{-1}\left(\sqrt{\frac{z}{z_0}}\right),\nonumber\\
A_t(z)&=& \mu+\rho  \left(\frac{z_0
z^2}{2}+\frac{z^3}{3}\right).\eea where $z_0,\mu,$ and  $\rho$ are
integration constants and $V_{21}$ is a constant from the dilaton
potential and $g_g$ is gauge coupling.  The dilaton potential is
given as \bea\label{dilatonpotential2}
V_E(\phi)&=&-\frac{12}{L^2}-\frac{42 \sinh ^4\left(\frac{\phi }{3
\sqrt{2}}\right)}{L^2}-\frac{42 \sinh ^2\left(\frac{\phi }{3
\sqrt{2}}\right)}{L^2}\nonumber\\&{}&-\frac{3 V_{21} \sinh
^{14}\left(\frac{\phi }{3 \sqrt{2}}\right)}{35 L^2}-\frac{3 V_{21}
\sinh ^{12}\left(\frac{\phi }{3 \sqrt{2}}\right)}{10
L^2}-\frac{3V_{21} \sinh ^{10}\left(\frac{\phi }{3
\sqrt{2}}\right)}{10
L^2}\nonumber\\
&{}&+\frac{\rho^2 z_0^8}{ g_g^2 L^2}\Big\{\frac{ \sinh
^{24}\left(\frac{\phi }{3 \sqrt{2}}\right)}{20  L^2}+\frac{3 \sinh
^{22}\left(\frac{\phi }{3 \sqrt{2}}\right)}{10 L^2}+\frac{59 \sinh
^{20}\left(\frac{\phi }{3
\sqrt{2}}\right)}{80L^2}\nonumber\\&{}&+\frac{15  \sinh
^{18}\left(\frac{\phi }{3 \sqrt{2}}\right)}{16 L^2}+\frac{5 \sinh
^{16}\left(\frac{\phi }{3 \sqrt{2}}\right)}{8 L^2}+\frac{5 \sinh
^{14}\left(\frac{\phi }{3 \sqrt{2}}\right)}{28 L^2}\Big\}. \eea This
solution is also a generalized one given in \cite{Li:2011hp}.

In this paper, we have not discussed thermodynamical properties of
these two sets of black hole solutions and possible applications in
hQCD model from gauge/gravity duality.


\end{document}